\documentclass[12pt]{iopart}
\usepackage[english]{babel}
\usepackage{graphicx}
\usepackage{textcomp}
\usepackage{cite}
\usepackage{color}

\newcommand{\w}{\phantom{1}}

\begin{document}

\title[Energy dependence of $pp\to\{pp\}_s\gamma$ in the $\Delta(1232)$ region]{Energy dependence of
hard bremsstrahlung production in proton-proton collisions in the
$\Delta$(1232) region}
\author{
D.~Tsirkov$^1$,
V.~Komarov$^1$,
T.~Azaryan$^1$,
D.~Chiladze$^{2,3}$
S.~Dymov$^{1,4}$,
A.~Dzyuba$^5$,
M.~Hartmann$^2$,
A.~Kacharava$^2$,
A.~Khoukaz$^6$,
A.~Kulikov$^1$,
V.~Kurbatov$^1$,
G.~Macharashvili$^{1,3}$,
S.~Merzliakov$^{1,2}$,
M.~Mielke$^{6}$,
S.~Mikirtychiants$^{2,5}$,
M.~Nekipelov$^{2}$,
F.~Rathmann$^2$,
V.~Serdyuk$^{1,2}$,
H.~Str\"oher$^2$,
Yu.~Uzikov$^1$,
Yu.~Valdau$^{2,5}$
and C.~Wilkin$^7$
}
\address{$^1$ Laboratory of Nuclear Problems, Joint Institute for Nuclear
Research, RU-141980 Dubna, Russia\\ %
$^2$ Institut f\"ur Kernphysik and J\"ulich Centre for Hadron
Physics, Forschungszentrum J\"ulich, D-52425 J\"ulich, Germany\\ %
$^3$ High Energy Physics Institute, Tbilisi State University,
0186 Tbilisi, Georgia\\ %
$^4$ Physikalisches Institut II, Universit{\"a}t
Erlangen-N{\"u}rnberg, D-91058 Erlangen, Germany \\ %
$^5$ High Energy Physics Department, Petersburg Nuclear
Physics Institute, RU-188350 Gatchina, Russia\\ %
$^6$ Institut f\"ur Kernphysik, Universit\"at M\"unster,
D-48149 M\"unster, Germany\\
$^7$ Physics and Astronomy Department, UCL, London, WC1E 6BT, United
Kingdom\ead{cw@hep.ucl.ac.uk}}

\date{\today}

\begin{abstract}
Hard bremsstrahlung production in proton-proton collisions has been
studied with the ANKE spectrometer at COSY-J\"ulich in the energy
range of 353--800~MeV by detecting the final proton pair
$\{pp\}_{\!s}$ from the $pp\to\{pp\}_{\!s}\gamma$ reaction with very
low excitation energy. Differential cross sections were measured at
small diproton c.m.\ angles from $0^{\circ}$ to $20^{\circ}$ and the
average over this angular interval reveals a broad peak at a beam
energy around 650~MeV with a FWHM $\approx$ 220~MeV, suggesting the
influence of $\Delta(1232)N$ intermediate states. Comparison with
deuteron photodisintegration shows that the cross section for
diproton production is up to two orders of magnitude smaller, due
largely to differences in the selection rules.
\end{abstract}


\submitto{\JPG}

\maketitle
%
%
\section{Introduction}

The photodisintegration of the deuteron%
\begin{equation}
\gamma d\to p n,\label{dgamma2pn}
\end{equation}
is the simplest reaction of its type and has therefore attracted much
attention, both theoretical and experimental, to learn more about how
to treat the interaction of the electromagnetic field with nuclei.
Significant advances in the understanding of the dynamics at several
hundred MeV have been achieved within the framework of models that
are either perturbative, taking into account a set of relevant
Feynman diagrams~\cite{LAG1989}, or employ a coupled $NN$, $N\Delta$,
$\Delta\Delta$ and $NN\gamma$ channel
formalism~\cite{ARE2003,LEI1987}. Nevertheless, extensive work on the
reaction continues and a full interpretation of the data has still to
be achieved~\cite{GLI2010}.

A kinematically similar reaction involves the photodisintegration of
a $^{1\!}S_0$ proton pair (diproton), here denoted by $\{pp\}_{\!s}$,
\begin{equation}
\gamma\{pp\}_{\!s}\to pp\,.
\label{ppgamma2pp}
\end{equation}
Reactions \eref{dgamma2pn} and \eref{ppgamma2pp} share the common
feature that, in the $\Delta(1232)$ range and above, there is a large
energy transfer from the photon to the target nucleon pair,
transforming it into one of high invariant mass and driving the final
state deep into the nucleon resonance region~\cite{GIL2002}. Thus, in
contrast to other electromagnetic processes, such as
electron-deuteron elastic scattering, the nucleon resonance
excitation channels are explicitly open. As a consequence, the
influence of isobar or mesonic exchange currents is likely to be very
strong for both reactions.

Since the diproton is the spin-isospin partner of the deuteron,
reactions \eref{dgamma2pn} and \eref{ppgamma2pp} involve different
transitions. Whereas one of the main driving terms for $\gamma d\to p
n$ is an $M1$ excitation of an $S$-wave $\Delta(1232)N$ pair that
de-excites into $pn$, an analogous transition is forbidden by angular
momentum and parity conservation for the $\gamma\{pp\}_{\!s}\to pp$
reaction~\cite{LAG1989}. There are likely to be cancellations among
the large amplitudes as one approaches the pure $S$-wave diproton
limit and this means that extra insight may be gained into the field
through a combined study of the photodisintegration of the deuteron
and diproton.

However, the free $^{1\!}S_0$ proton pair is not bound and the
diproton investigation has generally been approached through the
photodisintegration of a $pp$ pair embedded in a light nucleus, in
particular $^3\mathrm{He}$. By studying events where there were two
fast protons emerging from the $\gamma^{3}\textrm{He}\to ppn$
reaction with a slow (reconstructed) neutron, this part of phase
space was primarily interpreted in terms of an interaction on a
diproton, with the neutron merely appearing as a \textit{spectator}
particle, whose influence is only felt through the
kinematics~\cite{AUD1989,HOS1989,AUD1991,AUD1993,SAR1993,EMU1994,TED1994,NIC2004,POM2010}.
These data show little indication for the excitation of an
intermediate $\Delta(1232)N$ state, certainly much less than for
those for fast $pn$ pairs~\cite{EMU1994,TED1994}.

As pointed out in several of these publications, the corrections to
the simple spectator model picture can be quite large, owing mainly
to the fact that the photoabsorption on $pn$ pairs is much stronger
than on $pp$. This makes it harder to separate cleanly these two
terms, especially at the lower energies where the Dalitz plot is not
sufficiently wide. Final state interactions, where the neutron from a
fast $pn$ pair undergoes a charge-exchange reaction on a spectator
proton, and other three-body mechanisms have also to be considered.

In view of these potential drawbacks, it is well worth seeing if the
information derived from these experiments could be complemented by a
direct measurement of the cross section for
\begin{equation}
pp\to\{pp\}_{\!s}\gamma\,. \label{ppgamma}
\end{equation}
In order to ensure that the final diproton is almost exclusively in
the $^{1\!}S_0$ state, it is crucial that the $pp$ excitation energy
$E_{pp}$ be very small, and this is an important constraint on any
measurement. For small $E_{pp}$, the proton beam energy $T_p$ is
essentially twice the photon energy $E_{\gamma}$ for the inverse
reaction.

There have been many measurements of bremsstrahlung in proton-proton
scattering in the few hundred MeV
range~\cite{NEF1979,MIC1990,PRZ1992,YAS1999,HUI2002,MAH2004}. These
were generally accomplished through the use of pairs of counters
placed on either side of the primary beam direction. As a consequence
they were not sensitive to the behaviour at low $E_{pp}$, i.e.\ to
the hardest part of the bremsstrahlung spectrum. The one early
exception benefitted from the wide acceptance offered by the COSY-TOF
spectrometer, where data were taken at 293~MeV~\cite{BIL1998}.
However, the statistics in the low $E_{pp}$ region were severely
limited and a cross section for this region was not given.

The first data on hard bremsstrahlung production in
reaction~\eref{ppgamma} at intermediate energies were reported
recently from an experiment carried out at the ANKE facility at the
Cooler Synchrotron COSY-J\"ulich~\cite{KOM2008}. The $pp\to
\{pp\}_{\!s}\gamma$ differential cross section was measured at
energies $T_p = 353$, 500 and 550~MeV. Events were here selected with
a final excitation energy $E_{pp}<3$~MeV, where it is expected that
the contribution from $P$-waves in the diproton should be minimised.
However, due to the limited angular coverage of the forward detector
employed in these studies, only small diproton c.m.\ angles
$\theta_{pp}$ were covered; $0^{\circ} <\theta_{pp}< 20^{\circ}$.
Since the cross section is symmetric about $90^{\circ}$, this
effectively means a similar cut on the photon angle
$\theta_{\gamma}$; $0^{\circ} <\theta_{\gamma}< 20^{\circ}$. In
contrast, the full $pp\to \{pp\}_{\!s}\gamma$ angular domain was
measured in a high statistics experiment at CELSIUS with the same
$E_{pp}$ cut~\cite{JOH2009}, though only at the single (lower) energy
of 310~MeV.

Taken together, the ANKE and CELSIUS data show a rise in the
near-forward cross section from 310 to 550~MeV, but data at higher
energy are needed to see if there is a maximum in the region of the
$\Delta(1232)N$ threshold. To clarify the experimental situation, we
have supplemented the earlier COSY-ANKE data~\cite{KOM2008} by
measurements of the small angle $pp\to \{pp\}_{\!s}\gamma$
differential cross section at $T_p = 625$, 700 and 800~MeV. Combining
the data sets allows us to study the energy dependence of the
near-forward cross section throughout the $\Delta$(1232) resonance
region. Unlike some of the $^3$He photodisintegration
results~\cite{EMU1994,TED1994}, our data reveal a significant maximum
in a region where $\Delta(1232)N$ intermediate states might be
expected to play a role and this should provide a useful guide for
further theoretical work.

In view of earlier publications, the description of the experiment
and its analysis in section~\ref{MandA} could be made more brief. The
results and their significance are discussed in
section~\ref{Results}, with our conclusions and suggestions for
further work being presented in section~\ref{Conclusions}.

%
%

\section{Measurement and Analysis}
\label{MandA}

The experiment was carried out using the magnetic spectrometer
ANKE~\cite{BAR2001}, which is installed at an internal target station
of COSY. A hydrogen cluster-jet target was positioned in the proton
beam and the secondary particles were detected with wire chambers and
a scintillation hodoscope. The three-momenta and trajectories of the
particles were reconstructed on the basis of the known field map of
the analysing magnet, assuming that these particles originated from a
point-like source situated at the centre of the target-beam
interaction volume.

\begin{figure}[htb]
\begin{center}
\includegraphics[width=0.67\textwidth]{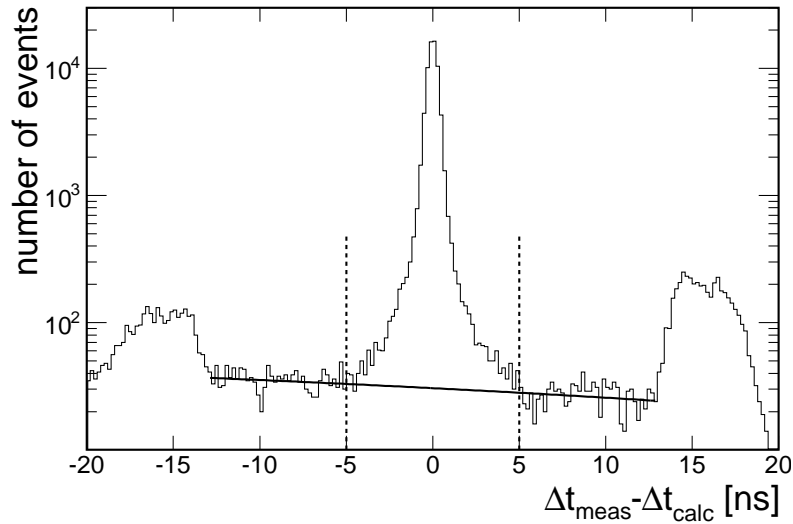}
\caption{Identification of the proton pairs from the $pp\to ppX$
reaction on the basis of time-of-flight information at 700~MeV.
$\Delta t_\mathrm{meas}$ is the directly measured difference of the
times of flight; $\Delta t_\mathrm{calc}$ is the time-of-flight
difference calculated using the measured particle momenta and
trajectories. The central peak corresponds to proton pairs, whereas
those around $\pm15$~ns are associated with $p\pi^+$ pairs. The full
line is a linear approximation to the background arising from
accidental coincidences. Events inside the indicated $\pm5$~ns
interval were retained for the analysis. } \label{tof}
\end{center}
\end{figure}

The first step in the identification of the $pp\gamma$ final state
was the selection of two coincident protons from among all the
detected pairs of positively charged particles. The scintillation
hodoscope allowed the measurement of the difference between the times
of flight from the target to the detector for the two recorded
particles. The comparison of this value with that \textit{calculated}
from the measured particle momenta and trajectories led to a very
good identification of proton pairs, as illustrated in
figure~\ref{tof}. The background from misidentified pairs was at the
few percent level for all beam energies, as detailed in
Table~\ref{lum}. Having identified two protons and determined their
momenta, the complete kinematics of the $pp\to ppX$ process could be
reconstructed. Events with zero missing mass, within the experimental
resolution, were accepted as candidates for the $pp\to pp\gamma$
reaction. Those where the pair's kinetic energy $E_{pp}$ in their
rest frame is small, specifically $E_{pp}<3$~MeV, were classified as
belonging to reaction~\eref{ppgamma}.

The experimental techniques, conditions of the measurements, and the
data-handling procedure have been described in an earlier
publication~\cite{KOM2008} and initial results given there at 353,
500, and 550~MeV. The new results at energies of 625, 700 and 800~MeV
were obtained in three separate beam periods. As a consequence, there
were some differences in the adjustment of the magnetic system setup
and the number of sensitive planes in the tracking detector. The 625
and 800~MeV data were collected as by-products of the study of other
reactions so that the measurement conditions were not optimised for
the study of the $pp\to pp\gamma$ reaction. The data at 700~MeV were
obtained with a polarised proton beam and a weighted average was then
taken over the two polarisation states.

\begin{table}[!ht]
\caption{\label{lum}Characteristics of the measurements at different
energies: integral luminosity $L$ with systematic (first) and
normalisation (second) errors; background/signal ratio
$N_\mathrm{bg}/N_{pp}$ for proton pair identification; width
FWHM($M_{\pi}^2$) of the $\pi^0$ peak in the missing-mass-squared
distribution from the $pp\to\{pp\}_{\!s}X$ reaction. The reasons for
the broader widths at 625 and 800~MeV are discussed in the text.}
\begin{indented}
\item[]\begin{tabular}{@{}cccc} \br %
$T_p$ & $L$ & $N_\mathrm{bg}/N_{pp}$ & FWHM($M_\pi^2$)$\times10^{-3}$ \\
(MeV) & ($10^{31}$cm$^{-2}$) &  & ((GeV/c$^{2}$)$^{2}$) \\ \mr
353 & $573\pm18\pm17$ & \phantom{1.}4\% & \w 4.1 \\
500 & $331\pm10\pm13$ & \phantom{1.}5\% & \w 6.3 \\
550 & $318\pm21\pm13$ & \phantom{1.}4\% & \w 7.4 \\
625 & $46\pm1\pm2$    &           1.6\% &   15.8 \\
700 & $159\pm3\pm8$\w &           2.2\% & \w 9.3 \\
800 & $67\pm1\pm3$    &           1.6\% &   18.9 \\ \br
\end{tabular}
\end{indented}
\end{table}

In order to measure the excitation energy of the proton pair and
study the angular dependence of the cross section of the $pp\to
pp\gamma$ reaction, sufficient resolution in the corresponding
variables is needed. The uncertainty $\sigma(\theta_{pp})$ in the
polar angle $\theta_{pp}$ of the diproton in the overall c.m.\ system
ranged from $0.5^{\circ}$ to $2.3^{\circ}$, depending on the beam energy and
the value of $\theta_{pp}$. The uncertainty in the excitation energy
$E_{pp}$, which generally increased with $E_{pp}$, was between 0.08
to 0.6~MeV for $E_{pp}< 3$~MeV, depending also on the beam energy.
The $E_{pp}$ resolution was thus sufficient for the measurement of
the excitation energy spectra up to 3~MeV. The $E_{pp}$ spectra for
$pp\to\{pp\}_{\!s}\gamma$, as well as for the
$pp\to\{pp\}_{\!s}\pi^0$ reaction, are satisfactorily reproduced by
Monte Carlo simulations, where events were generated according to
phase space modified by the $S$-wave $pp$ final state
interaction~\cite{KOM2008,DYM2006}. Extra evidence for the $S$-wave
nature of the proton pairs was provided by the isotropy of the
acceptance-corrected angular distributions in the diproton rest
frame.

\begin{figure}[htb]
\begin{center}
\includegraphics[width=1\textwidth]{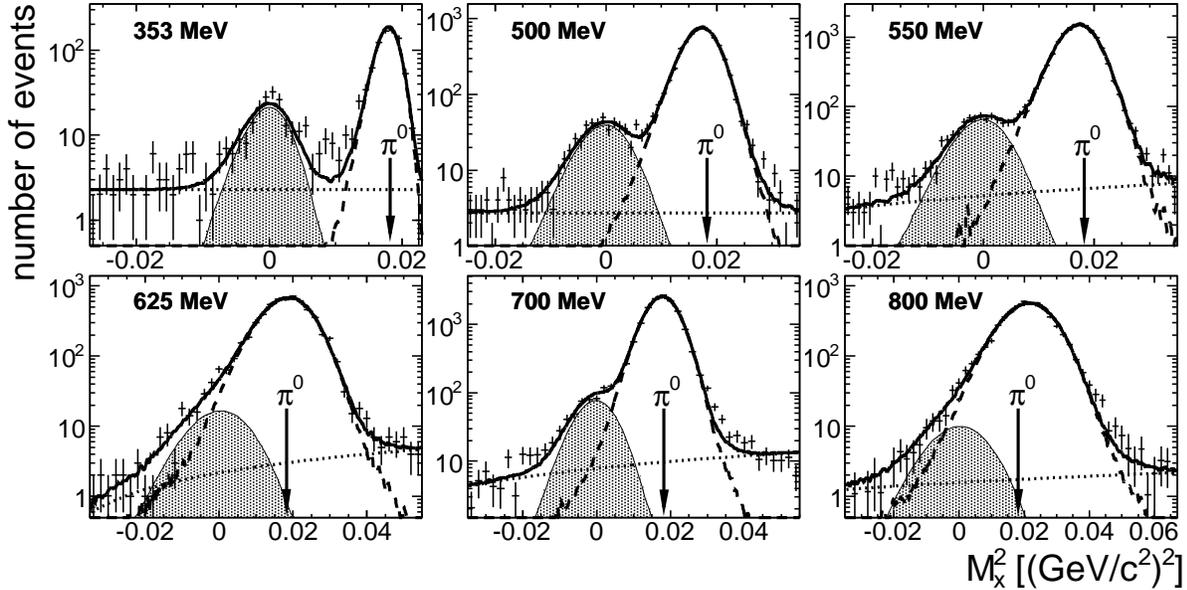}
\caption{Distributions of the missing mass squared in the
$pp\to\{pp\}_{\!s}X$ reaction for proton pairs with
$\theta_{pp}<20^{\circ}$ compared to fits with different
contributions. The expected $\pi^0$ position is indicated by the
arrow. The shaded area corresponds to the $\gamma$ peak, the dashed
line to the $\pi^0$ peak, the dotted to the linear accidental
background, and the solid to the sum of these three contributions.
The data at the three lower energies were reported in \cite{KOM2008},
whereas the others are from this work.} \label{Mx2fit}
\end{center}
\end{figure}

The distributions in missing mass squared $M_x^2$ for events where
$\theta_{pp}< 20^{\circ}$ are shown in figure~\ref{Mx2fit} for the
six beam energies. At 353~MeV there is a clear $\gamma$ peak that is
well separated from the pion peak associated with the
$pp\to\{pp\}_{\!s}\pi^0$ reaction. The low background arising from
accidental proton coincidences changes weakly with $M_x^2$ and could
be taken as linear in the interval $-0.02 < M_x^2
<+0.06~($GeV/c$^2)^2$.

The widths of the peaks in figure~\ref{Mx2fit} are governed by the
accuracy of the measurement of the proton three-momenta in the
relevant experimental runs and this generally gets worse with
increasing beam energy. The corresponding growth in the widths
reported in Table~\ref{lum} results at higher energies in the merging
of the $\gamma$ signal with that of the pion. The principal
difficulty consists therefore in selecting the small number of
$\gamma$ events from the total distribution.

The $M_x^2$ distributions were fitted by a sum of peaks,
corresponding to $\gamma$ and $\pi^0$ production, plus a linear
background. In order to estimate the shapes of the peaks as reliably
as possible, a detailed Monte-Carlo simulation was undertaken at each
energy, taking into account all the known features of the setup.
These include, in particular, the smearing caused by the radial
distribution of the proton beam at the target region, the multiple
scattering in the exit window of the vacuum chamber and detector
materials, and the actual clustering of the wires fired in the
proportional chambers. The procedure of track reconstruction used in
the simulation was the same as that in the data handling. The free
parameters of interest used to fit the missing-mass spectra were the
number of events in the $\gamma$ peak, the number of events in the
pion peak, and the two constants of the linear background. However,
in order to compensate for the lack of knowledge of the beam spatial
distribution, additional parameters were inserted into the fits.
These were the shift of the pion peak position and correction factors
for the $\gamma$ and pion peak widths.

In order to investigate the systematics, several fitting methods were
employed. Either both peaks were fitted together or the pion peak
with the background was first fitted separately, excluding the
$M_x^2$ range where the $\gamma$ signal was expected. The correction
for the setup acceptance was also accounted for in two different
ways. This was either introduced for each event and the weighted
$M_x^2$ spectrum fitted, or the uncorrected distribution was fitted
and the number of $\gamma$ events corrected for the average
acceptance factor afterwards. Finally, two fitting approaches were
tried, namely $\chi^2$ minimisation and logarithmic likelihood,
though the latter was used only for fitting uncorrected spectra.
Although these techniques were developed primarily in order to
identify the $\gamma$ signal in the more difficult high energy cases,
they were also used on the published data~\cite{KOM2008}, but any
changes there are of little significance.

\begin{figure}[!ht]
\begin{center}
\includegraphics[width=1\textwidth]{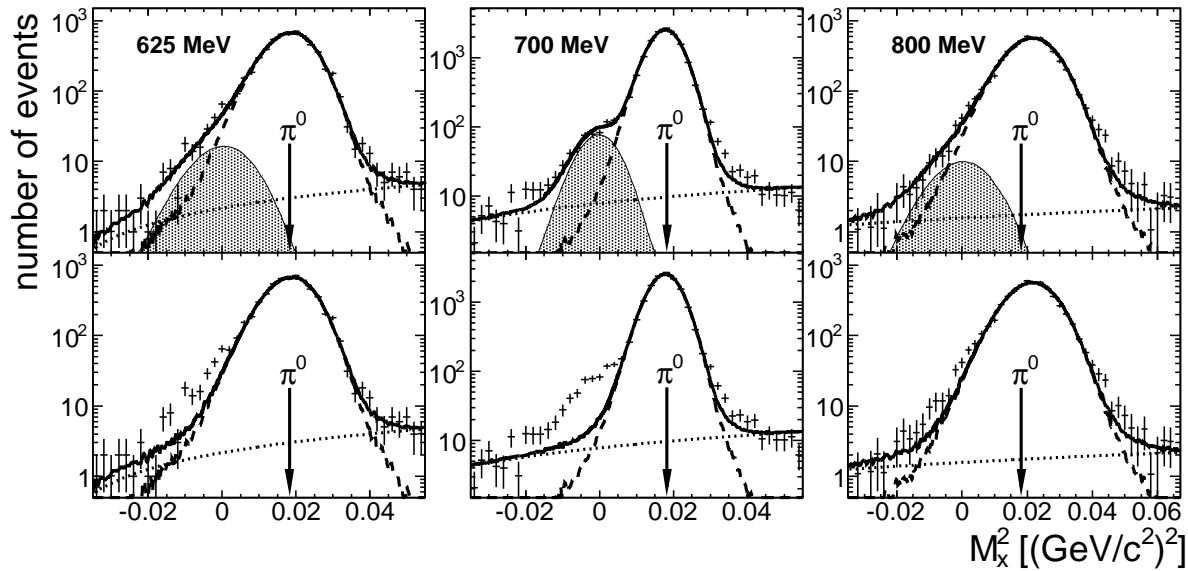}
\caption{Descriptions of the $M_x^2$ distributions at 625, 700 and
800~MeV, assuming the presence (upper panels) or absence (lower
panels) of a $\gamma$ contribution, as explained in the text. The
notation is as in figure~\ref{Mx2fit}. The values $\chi^2$ per degree
of freedom in the $\gamma$ peak region increase significantly for
fits that do not include the possibility of a $\gamma$ signal. At
625~MeV the $\chi^2/{\rm ndf}$ goes from 20.8/9 to 69.8/10, at
700~MeV from 9.7/9 to 314/10, and at 800~MeV from 9.7/9 to 36/10.}
\label{mx2chi2}
\end{center}
\end{figure}

At each energy the results obtained using the different methods were
completely compatible and their average was taken, with the variation
in the deduced cross section being considered as a systematic
uncertainty. As a further check on the reliability of the
identification of the $\gamma$ signal in the most complicated cases
of higher $T_p$, the $M_x^2$ distributions were also fitted assuming
that there were no $\gamma$ events. In this case the $\pi^0$ peak was
first fitted, excluding the $\gamma$ range around $M_x^2 =0$, and the
$\chi^2$ of the difference between the resulting function and the
histogram evaluated over the previously excluded $\gamma$ range. The
results of fitting the data at 625, 700 and 800~MeV with and without
the $\gamma$ contribution are shown in figure~\ref{mx2chi2}, where it
is seen that, unless the $\gamma$ is included, there is a significant
deterioration in the values of $\chi^2$ evaluated for the $\gamma$
peak region even in the most severe case of 800~MeV.

The luminosity $L$ was determined from the number of the elastically
scattered protons detected in parallel. In order to correct for the
acceptance, a simulation was carried out where the setup geometry,
the efficiency of the proton detection by the multiwire proportional
chambers, and the track reconstruction algorithm were taken into
account. These measurements were then compared to differential cross
sections predicted by the SAID analysis program~\cite{ARN2007}.
Although this does not yield errors, study of experimental data in
the literature suggests that the uncertainties range from about 3\%
at 353~MeV up to 5\% at 800~MeV. Small variations of the luminosity
(those exceeding the statistical fluctuations), derived at different
proton angles, reflect uncertainty in the acceptance correction and
were taken as systematic errors in the luminosity.

The total uncertainty was estimated as the quadratic sum of the
statistical, systematic, and normalisation errors. As a general rule,
statistical errors had a negligible effect so that they are not shown
in Table~\ref{lum}. Also negligible is the contribution arising from
the systematic error in the proton momentum.

The angular dependences at 353, 500 and 550~MeV were reported
earlier~\cite{KOM2008}. The poorer conditions did not allow similar
studies to be made at 625 and 800~MeV. However, to estimate the
angular dependence of the differential cross section at 700~MeV, we
have divided the events into three $\theta_{pp}$ intervals,
$0^{\circ}$--$7^{\circ}$, $7^{\circ}$--$12^{\circ}$,
$12^{\circ}$--$20^{\circ}$, and made separate fits for each of these
ranges.

%
%
\section{Results and Discussion}
\label{Results}

In order to study the energy dependence of the $pp\to\{pp\}_{\!s}\gamma$
reaction from 353 to 800~MeV, we have evaluated the differential
cross section $\rmd\sigma/\rmd\Omega_{(0-20)}$ averaged over the
interval $0^{\circ}-20^{\circ}$ in diproton c.m.\ angle. Our
measurements show that the angular dependence is rather smooth in
this region~\cite{KOM2008} so that the average cross section should
reflect reasonably well the energy dependence at fixed angle.

The numbers of selected $\gamma$ events, $N_\mathrm{raw}$, and the
corresponding acceptance-corrected figure, $N_\mathrm{corr}$, are
given in Table~\ref{tabdserr} for the six energies. Also to be found
there are the average cross sections
$\rmd\sigma/\rmd\Omega_{(0-20)}$, together with the corresponding
partial and total errors. The uncertainties are particularly large at
625 and 800~MeV, due to the low luminosity and non-optimal conditions
that hampered the identification of the $\gamma$ signal.

\begin{table}[!ht]
\caption{\label{tabdserr}Numbers of $pp\to\{pp\}_{\!s}\gamma$ events
and the corresponding differential cross sections at different beam
energies. Data at 353, 500, and 550~MeV were already presented
in~\cite{KOM2008}. Here $N_\mathrm{raw}$ is the number of registered
$\{pp\}_{\!s}\gamma$ events, $N_\mathrm{corr}$ the number of events
corrected for acceptance, and $\rmd\sigma/\rmd\Omega_{(0-20)}$ the
differential cross section averaged over $0^{\circ}-20^{\circ}$. The
statistical error is denoted by $\sigma_\mathrm{stat}$, that coming
from the systematics in the $\gamma$ event acquisition by
$\sigma_\mathrm{\gamma}$, and that arising from the luminosity
uncertainty by $\sigma_\mathrm{lum}$. Adding these contributions
quadratically gives a total error of $\sigma_\mathrm{tot}$.}
\begin{indented}
\item[]\begin{tabular}{@{}ccccccc} \br %
$T_p$ (MeV)                         &    353 &    500 &    550 &    625 &    700 &    800 \\ \mr
$N_\mathrm{raw}$                    &    180 &    335 &    525 &    177 &    450 &    114 \\
$N_\mathrm{corr}$                   & 1126\w & 2164\w & 3722\w &    810 & 2296\w &    459 \\ \mr
$\rmd\sigma/\rmd\Omega_{(0-20)}$ (nb/sr) & 5.1 & 17.9 &   33.5 &   46.5 &   37.9 &   17.0 \\
$\sigma_\mathrm{stat}$, nb/sr       &   0.4 &  \w 1.0 & \w 1.4 & \w 6.4 & \w 2.6 & \w 4.9 \\
$\sigma_\mathrm{\gamma}$, nb/sr     &   0.2 &  \w 0.5 & \w 0.7 & \w 7.0 & \w 1.0 & \w 8.5 \\
$\sigma_\mathrm{lum}$, nb/sr        &   0.2 &  \w 0.9 & \w 2.6 & \w 2.0 & \w 2.2 & \w 1.0 \\
$\sigma_\mathrm{tot}$, nb/sr        &   0.5 &  \w 1.4 & \w 3.0 & \w 9.7 & \w 3.5 & \w 9.9 \\ \br
\end{tabular}
\end{indented}
\end{table}

\begin{figure}[!ht]
\begin{center}
\includegraphics[width=0.67\textwidth]{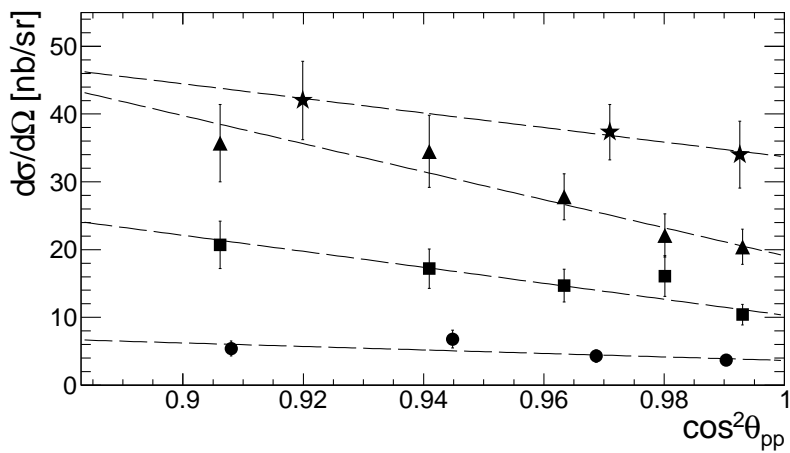}
\caption{Angular dependence of the differential cross section for the
$pp\to\{pp\}_{\!s}\gamma$ reaction at four beam energies. The results
of the present measurement at 700~MeV are shown by stars. Also shown
are the data from \cite{KOM2008} at 353~MeV (circles), 500~MeV
(squares), and 550~MeV (triangles). The errors coming from the
statistics and background subtraction are shown but not the overall
ones arising from the luminosity uncertainty. The lines represent
linear fits to the four data sets.} \label{dsdO}
\end{center}
\end{figure}

As previously remarked, it was also possible to extract information
on the angular dependence of the $pp\to\{pp\}_{\!s}\gamma$ reaction
at 700~MeV by dividing the ANKE range into three intervals and the
results are presented in Table~\ref{tabdsdO}. These data are compared
in figure~\ref{dsdO} with those obtained at lower
energies~\cite{KOM2008}. Although the new data show some tendency for
a forward dip, this is less strong than that measured at 500 and
550~MeV. In all cases, over the limited angular interval reported
here, the cross section is consistent with a linear variation in
$\cos^2\theta_{pp}$.

\begin{table}[!ht]
\caption{\label{tabdsdO}Angular dependence of the
$pp\to\{pp\}_{\!s}\gamma$ reaction at 700~MeV.}
\begin{indented}
\item[]\begin{tabular}{@{}cccc} \br %
$\theta_{pp}$ & $0^{\circ}$--$7^{\circ}$ & $7^{\circ}$--$12^{\circ}$
& $12^{\circ}$--$20^{\circ}$ \\ \mr
$\rmd\sigma/\rmd\Omega$ (nb/sr) & 34$\pm$5 & 37$\pm$4 & 42$\pm$6 \\
\br
\end{tabular}
\end{indented}
\end{table}

\begin{figure}[!ht]
\begin{center}
\includegraphics[width=0.67\textwidth]{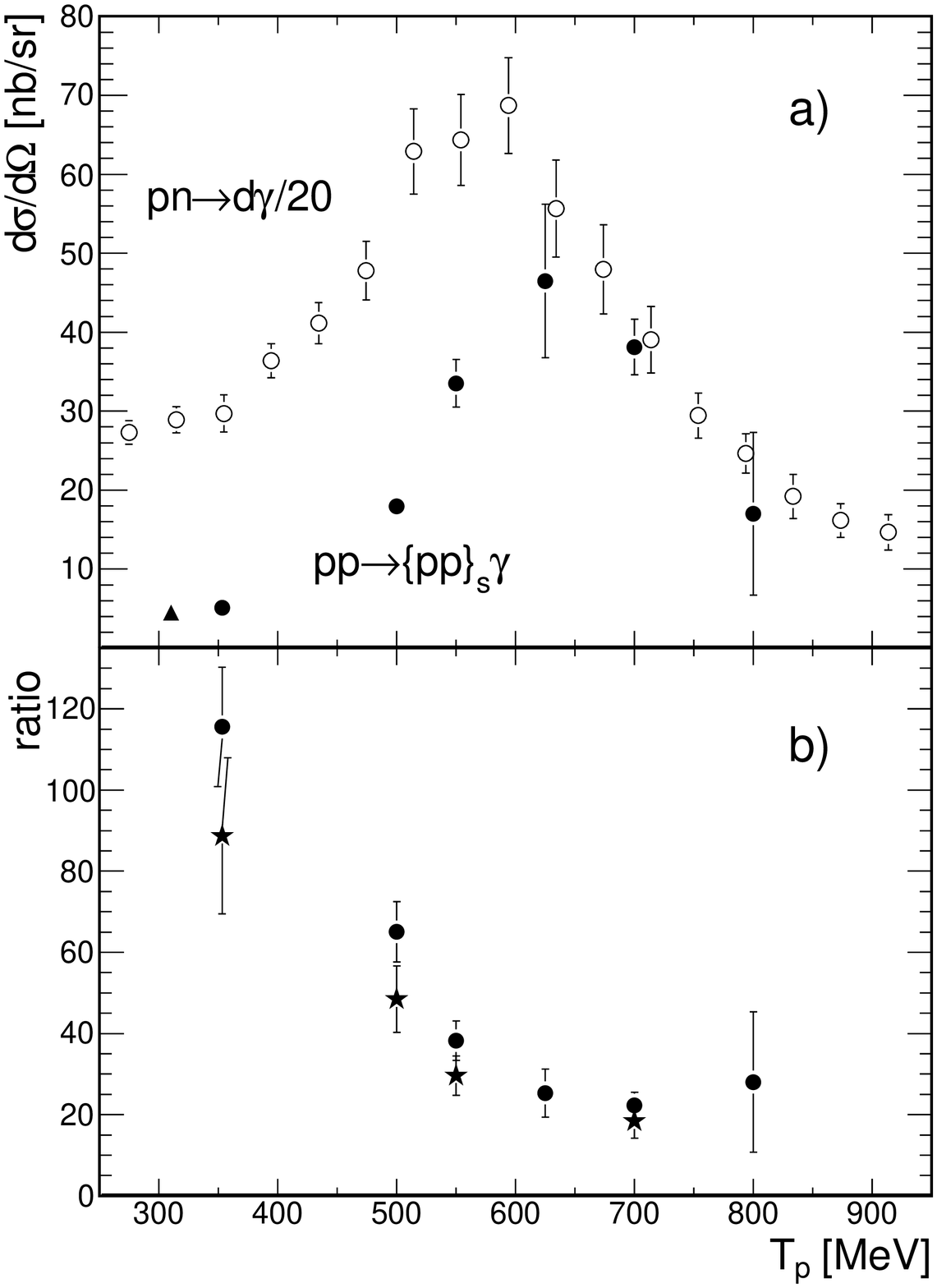}
\caption{(a) Energy dependence of the $pp\to\{pp\}_{\!s}\gamma$ and
$pn\to d\gamma$ differential cross sections. Full circles represent
COSY-ANKE data for the average differential cross section
$\rmd\sigma/\rmd\Omega_{(0-20)}$ for the $pp\to\{pp\}_{\!s}\gamma$
reaction whereas the triangle is taken from the CELSIUS
results~\cite{JOH2009} for the same angular and $E_{pp}$ conditions.
The open circles show the values of the differential cross section
for the $pn\to d\gamma$ reaction at a deuteron c.m.\ angle of
$\theta_d=20^{\circ}$ with respect to the proton direction, evaluated
from the deuteron photodisintegration data of \cite{CRA1996} and
scaled down by a factor of twenty. (b) Ratio of the differential
cross sections for $pn\to d\gamma$ to $pp\to\{pp\}_{\!s}\gamma$, as
defined by equation~\eref{eqratio}. The stars were obtained for the
same production angle $\theta_d=\theta_{pp}=20^\circ$ while for the
circles the angular average of the $pp\to\{pp\}_{\!s}\gamma$ cross
section over $0^{\circ}-20^{\circ}$ was used.} \label{dsdE}
\end{center}
\end{figure}

The measured values of the average cross section
$\rmd\sigma/\rmd\Omega_{(0-20)}$ are shown in figure~\ref{dsdE}a as a
function of the proton beam energy. Similar data are also available
with the identical 3~MeV cut in $E_{pp}$ at 310~MeV from
CELSIUS~\cite{JOH2009}. Although this experiment had a much wider
angular coverage, the value shown in the figure was obtained by
taking the $0^{\circ}-20^{\circ}$ average. This point joins very
smoothly onto the ANKE data.

The energy dependence of the differential cross section of the
$pp\to\{pp\}_{\!s}\gamma$ reaction shown in figure~\ref{dsdE}a
reveals a broad peak, with a maximum near 650~MeV and FWHM $\sim
220$~MeV, and it is tempting to suggest that this might be associated
in some way with the excitation of the $\Delta(1232)$ isobar. It is
well known that this resonance plays a crucial role for the analogous
$pn\to d\gamma$ reaction, whose energy dependence at a deuteron
production angle of $\theta_d=20^{\circ}$ with respect to the proton
direction is also shown in the same figure, scaled down by a factor
of twenty. These values were obtained from the MAMI deuteron
photodisintegration $\gamma d\to np$ results~\cite{CRA1996} by using
detailed balance.

It is important to note that the maximum in the
$pp\to\{pp\}_{\!s}\gamma$ cross section appears at about 80~MeV
higher than that for $pn\to d\gamma$. Direct excitation of the
$\Delta$-isobar dominates the $pn\to d \gamma$ reaction in this
energy range through an $M1$ transition to an intermediate $\Delta N$
state in an $S$-wave. Such an $M1$ transition is forbidden in the
diproton case~\cite{LAG1989} but the strength could come from a
$P$-wave $\Delta N$ configuration, which requires extra energy in
order to overcome the centrifugal barrier. Similar arguments have
been used to explain why the $\Delta$ peak is shifted to higher
energy also for the $pp\to\{pp\}_{\!s}\pi^0$ reaction~\cite{NIS2006}.
In addition to the displacement of the $pp\to\{pp\}_{\!s}\gamma$ peak
position to higher energies, it is narrowed significantly compared to
$pn\to d\gamma$.

The ratio of the two differential cross sections
\begin{equation}R=\frac{\rmd\sigma/\rmd\Omega(pn\to
d\gamma)}{\rmd\sigma/\rmd\Omega(pp\to\{pp\}_{\!s}\gamma)}\label{eqratio}\end{equation}
is presented in figure~\ref{dsdE}b. Where necessary, the $pn\to
d\gamma$ data~\cite{CRA1996} were spline-interpolated in energy. The
smallest c.m.\ angle in these measurements was $20^{\circ}$ and these
are compared in the figure to the diproton results both at this angle
and averaged between $0^{\circ}$ and $20^{\circ}$.

What is striking about the ratio shown in figure~\ref{dsdE}b is the
smallness of the diproton cross section compared to that for the
deuteron. The factor is strongly energy dependent, dropping smoothly
from over 100 at low energies to about 20 at 800~MeV. It should,
however, be borne in mind that some of this suppression is to be
associated with the difference in the phase space volume for
producing the deuteron bound state and diproton continuum state.

The theoretical treatment of the $pp\to\{pp\}_{\!s}\gamma$ reaction
has been far less developed than that for $pn\to d\gamma$. The first
thing to note is that, in the diproton case, the $E1$ transition is
suppressed by the vanishing of an electric dipole operator for the
proton pair. The absence of charged pion exchange currents also
reduces some of the meson exchange effects. Finally, because the
$S$-wave $J^P=1^+$ $\Delta N$ intermediate state cannot couple to an
initial $pp$ system~\cite{LAG1989}, the $M1$ transition that
dominates the $pn\to d\gamma$ reaction in the few hundred MeV region
is also not present. One might therefore expect the first effects of
the $\Delta$ degrees of freedom to show up in an $S$-wave $J^P=2^+$
$\Delta N$ contribution, which would lead to an $E2$ transition, or
through a $\Delta N$ configuration in a $P$ or higher wave. The
coupled-channel calculations of~\cite{WIL1995} suggest that at
intermediate energies the $E2$ multipole should dominate and this
would give rise to a $\sin^2\theta_{\gamma}\cos^2\theta_{\gamma}$
dependence. In addition, this model showed no sign for any apparent
$\Delta N$ signal in the energy variation, in contrast to our data.
Furthermore, the CELSIUS data at 310~MeV show a linear variation with
$\cos^2\theta_{\gamma}$ so that any $E2$ contribution at this energy
should be very small.

In a recent publication~\cite{NAK2009} the high precision KVI $pp$
bremsstrahlung data at 190~MeV~\cite{HUI2002} have been successfully
described for the first time through the introduction of a
phenomenological contact interaction current that explicitly
satisfies the generalised Ward-Takahishi identity. However, these
data did not sample the hard bremsstrahlung limit. A generalisation
of the theoretical model includes some $\Delta$ contributions, but it
still neglects entirely the interaction current due to the $\Delta$
that gives rise to the five-point contact current~\cite{NAK2009a}. It
is not clear if this is the reason why it did not reproduce the shape
of the angular distribution for small $E_{pp}$ from CELSIUS at
310~MeV~\cite{JOH2009}.

A peak in the energy dependence of the $pp\to\{pp\}_{\!s}\gamma$
differential cross section due to the $\Delta N$ configuration was
obtained in calculations~\cite{UZI2008} within the framework of a
simplified one-pion-exchange model. However, gauge invariance was not
imposed and other terms neglected.

%
%

\section{Summary and conclusions}
\label{Conclusions}

The differential cross section for hard bremsstrahlung production in
proton-proton collisions has been measured in the near-forward
direction, $0^{\circ}-20^{\circ}$, for beam energies between 353 and
800~MeV. The energy dependence of this cross section reveals a broad
peak around 650~MeV, which is roughly where one might expect a
contribution from a $\Delta(1232)N$ intermediate state. The peak is
shifted to higher energies compared to that in the analogous $pn\to
d\gamma$ reaction but this is not surprising because the $S$-wave
$J^p=1^+$ $\Delta N$ state that drives the $M1$ transition in the
deuteron case does not couple to an initial $pp$ system.

The suppression of the isobar $M1$ term leads to a large but
energy-dependent factor between the cross sections for $pn\to
d\gamma$ and $pp\to\{pp\}_{\!s}\gamma$. It also makes the theoretical
description harder to realise because there is then no longer an
obviously dominant term to consider. Further theoretical work is
clearly needed.

Even with a 3~MeV cut in $E_{pp}$ there might be some small
contamination of $P$-waves in the $pp$ system but there are similar
concerns for the $^3$He$(\gamma,pp)n$ data because variational Monte
Carlo calculations suggest that the $pp$ pair in $^3$He is not in a
pure $^{1\!}S_0$ state and might also contain a few per cent of
higher partial waves~\cite{WIR2008}.

The combined study of the $pn\to d\gamma$ and
$pp\to\{pp\}_{\!s}\gamma$ reactions has assumed a greater importance
in recent years because of the interest in the investigation of
short-range $pn$ and $pp$ correlations in
nuclei~\cite{POM2010,SHN2007,SUB2008}. These are studied through
photoabsorption that leads to the emission of nucleon pairs with
large back-to-back momenta in the pair's rest frame.

On the experimental side, the separation of the various multipoles in
the $pp\to \{pp\}_{\!s}\gamma$ reaction would clearly require the
measurement of the cross section (and analysing power) over a wider
angular region and this will be possible at ANKE through the use of a
positive side detector in combination with the forward
detector~\cite{DYM2008} that provided the data reported here. Of
great use in the separation would be data on the proton-proton spin
correlation parameters and it might be possible to study these at
COSY-ANKE through the use of a polarised gas cell
target~\cite{KAC2005}.

New experimental data~\cite{POM2010} has allowed the comparison of
the cross sections for $d(\gamma,p)n$ and $^3\mathrm{He}(\gamma,pp)n$
at $\theta_\mathrm{cm}=90^{\circ}$. These data come from the scaling
energy region, $E_\gamma> 2.4$~GeV, where the influence of the
additional nucleon in the $^3\mathrm{He}$ target is of less
importance. The ratio of the cross sections found there was
$\approx40$, but to compare with the $R$ ratio of equation
\ref{eqratio}, one needs to transform data with a bound diproton to
ones with a scattering $pp$ system, which requires a model
calculation.

An extension of the experimental study of the $pp\to
\{pp\}_{\!s}\gamma$ reaction from the $\Delta$(1232) excitation
region to GeV energies could provide an alternative way to
investigate the transition to a situation where hadronic internal
degrees of freedom determine the interaction. The onset of the QCD
scaling regime might take place in $pp\to \{pp\}_{\!s}\gamma$  at a
beam energy $T_p\approx 2$~GeV, corresponding to the $E_\gamma\approx
1$~GeV which, it is suggested, is the start of the domain for
deuteron photodisintegration~\cite{ROS2005}.

Although the maximum proton beam energy at COSY is nearly 3~GeV, the
identification of the $pp\to \{pp\}_{\!s}\gamma$ reaction through the
$pp$ missing mass becomes progressively harder as the energy
increases, as is well illustrated by the data in figure~\ref{Mx2fit}.
Therefore any attempt to use this reaction for the investigation of
the $\gamma NN$ dynamics and its relation to the quark degrees of
freedom would certainly necessitate the detection of the $\gamma$ in
coincidence. This is a challenge for the future.

\ack %
The authors wish to thank other members of the ANKE collaboration for
their help and assistance in the running of the experiment. We are
grateful also to the COSY crew for providing good working conditions.
Correspondence with Kanzo Nakayama has been very helpful. This work
has been partially supported by the BMBF (grant ANKE COSY-JINR), RFBR
(09-02-91332), DFG (436 RUS 113/965/0-1), the JCHP FFE, and the
HGF-VIQCD.
%
%

\end{document}